\documentclass[lettersize,journal]{IEEEtran}
\usepackage{amsmath,amsfonts}
\usepackage{algorithmic}
\usepackage{algorithm}
\usepackage{array}
\usepackage[caption=false,font=normalsize,labelfont=sf,textfont=sf]{subfig}
\usepackage{textcomp}
\usepackage{stfloats}
\usepackage{url}
\usepackage{verbatim}
\usepackage{graphicx}
\usepackage{cite}
\usepackage{color}
\begin{document}

\title{Sharing Intelligent Reflecting Surfaces in Multi-Operator Communication Systems for Sustainable 6G Networks}

\author{
    Hiroaki~Hashida,~\IEEEmembership{Member,~IEEE,}
    Yuichi~Kawamoto,~\IEEEmembership{Member,~IEEE,}
    and Nei~Kato,~\IEEEmembership{Fellow,~IEEE}
\thanks{H. Hashida is with Frontier Research Institute for Interdisciplinary Sciences, Tohoku University, Sendai, Japan, and H. Hashida, Y. Kawamoto, and N. Kato are with the Graduate School of Information Sciences (GSIS), Tohoku University, Sendai, Japan,
E-mail: \{hiroaki.hashida.d6, yuichi.kawamoto.d4, nei.kato.d3\}@tohoku.ac.jp.}
}

\markboth{IEEE Wireless Communications}{Draft}%

\maketitle

\begin{abstract}
In this study, we investigate the use of intelligent reflecting surfaces (IRSs) in multi-operator communication systems for 6G networks, focusing on sustainable and efficient resource management. 
This research is motivated by two critical challenges: limited coverage provided by mmWave frequencies and high infrastructure costs associated with current technologies. IRSs can help eliminate these issues because they can reflect electromagnetic waves to enhance signal propagation, thereby reducing blockages and extending network coverage.
However, deploying a separate IRS for each mobile network operator (MNO) can result in inefficiencies, redundant infrastructure, potential conflicts over placement, and interoperator interference.
To address these challenges, in this study, an IRS sharing system is proposed in which multiple MNOs collaborate to use a common IRS infrastructure. This approach not only enhances network flexibility and reduces costs but also minimizes the effect of interoperator interference. Through numerical analysis, we demonstrate that IRS sharing effectively balances performance and fairness among MNOs, outperforming MNO-specific deployment methods in multi-MNO scenarios. This study provides insights into the potential of IRS sharing to support sustainable 6G networks, thereby contributing to the efficient deployment and operation of next-generation wireless communication systems.
\end{abstract}


\section{Introduction}
Massive multiple-input multiple-output (MIMO) and millimeter-wave (mmWave) communication are widely adopted in fifth-generation (5G) wireless networks to enhance spectral efficiency and network capacity, thereby supporting a greater number of users and advanced services~\cite{MIMO}. However, due to the high susceptibility of mmWave signals to blockage effects, propagation loss becomes significant~\cite{mina}. Hence, the coverage area of a single base station is severely limited, necessitating the deployment of numerous base stations, leading to elevated infrastructure costs. Although massive MIMO systems can partially mitigate these issues by enhancing link reliability and spatial diversity, their implementation incurs considerable energy consumption and infrastructure expenses. Moreover, while exploiting higher frequency bands—such as mmWave and terahertz waves—enables high-capacity communication, these bands inherently introduce additional propagation challenges. Consequently, sixth-generation (6G) wireless networks must develop innovative solutions to overcome these limitations and fully exploit the expanded bandwidth required for emerging applications~\cite{hashida_JSAC,6G_ansari}.
However, shifting to these higher frequency bands introduces additional complexity and costs, particularly in terms of infrastructure and power usage. Therefore, 6G should provide superior performance and facilitate sustainable and efficient resource management to satisfy the increasing demand for emerging applications~\cite{emerging_app,Ansari}.

An intelligent reflecting surface (IRS), also referred to as a reconfigurable intelligent surface (RIS), has emerged as a solution to address some inherent challenges of 6G networks, particularly in overcoming limitations related to higher-frequency bands such as mmWave and terahertz~\cite{hashida_TVT_selective}. This device comprises surfaces embedded with passive elements that can dynamically control the reflection characteristics of electromagnetic waves, thereby enabling efficient signal propagation. We can mitigate blockages, extend coverage, and reduce path loss by strategically placing the IRS on buildings or other structures and controlling its reflection characteristics, that is, phase shift, addressing some of the problems that have hindered the adoption of mmWave technology. Unlike conventional network components, IRSs do not require power-intensive transmitters; instead, they manipulate existing signals, which can considerably lower the operational costs and energy consumption~\cite{IRS_energy_efficiency,Zhu_JSAC}. 

The conventional approach to IRS deployment involves each mobile network operator (MNO) deploying and managing its own IRS to provide reliable connectivity in a region known as a single MNO-specific IRS system. However, as presented in Fig.~\ref{fig:motivation_of_IRS_sharing}, installing IRSs in networks in which multiple MNOs coexist presents the following challenges:

\begin{figure*}[t]
    \centering
    \includegraphics[width=1\hsize]{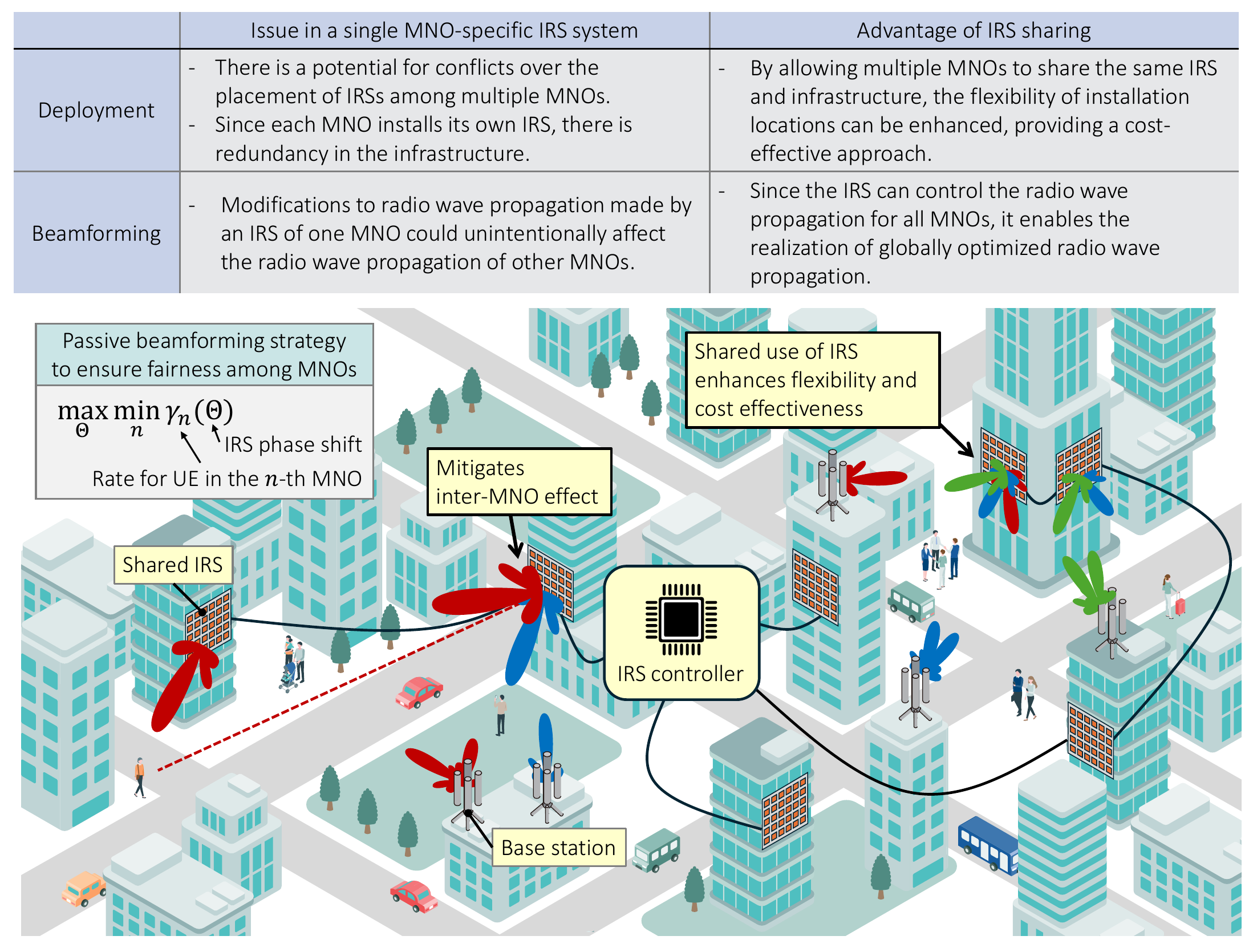}
    \caption{Motivation of IRS sharing.}
    \label{fig:motivation_of_IRS_sharing}
\end{figure*} 

\subsubsection{Conflicting IRS installation locations}
The passive nature of IRSs demands large surface areas—typically building façades or rooftops—to ensure sufficient aperture relative to the carrier wavelength and achieve adequate reflection gain. Moreover, studies have revealed that the placement of an IRS considerably affects its performance~\cite{Hashida_WCM}. Therefore, deploying a sufficiently sized IRS at an optimal location is necessary to leverage the IRS’s capabilities. However, in urban environments, such expansive installation locations are scarce due to space constraints and the complexities involved in obtaining permission to use privately owned buildings. Consequently, optimal sites become highly competitive, with multiple MNOs vying for the same prime locations. In the conventional business model, in which MNOs typically independently install communication infrastructure, there is often competition among MNOs for optimal placement locations.

\subsubsection{System Redundancy}
Despite sufficient installation space being available, the optimal placement of IRSs to eliminate areas that exhibit weak receiving power can be the same for each MNO. Consequently, multiple MNOs can deploy unique IRSs in the same area, increasing system redundancy. Moreover, when an IRS operates in conjunction with each MNO’s base station (BS), a control link is required between the BS and the IRS. Additionally, although IRSs are low-power devices, they require a small power supply to enable the switching of reflection properties and receiving of control signals, necessitating the provision of a power supply system. Furthermore, IRSs require large aperture areas, resulting in the IRS covering a considerable portion of the façade of a building. Such an increase in control equipment and wiring, as well as large-scale deployment, can increase energy consumption and affect the aesthetics of buildings and urban spaces. Therefore, deploying separate IRSs for each MNO can considerably render wireless network deployments unsustainable from a system deployment perspective.

\subsubsection{Inter-Operator Effects}
When IRS-aided cellular networks are deployed by various MNOs in overlapping geographical areas, the propagation characteristics can change unintentionally~\cite{pilot_contamination}. For instance, when an IRS owned by a specific MNO modifies the radio wave propagation between a BS and a UE, the propagation channels of other MNOs can be affected because even if IRS elements are designed for specific frequencies, they do not function as bandpass filters. Instead, they reflect broadband signals with uniform reflection coefficients across frequencies. Consequently, in communication systems that use passive IRSs, the IRS can unintentionally reflect signals across multiple bands, causing unpredictable and disruptive signals to other MNO's terminals. In MIMO communications, this phenomenon can lead to pilot contamination during channel estimation and intersymbol interference because of channel errors during demodulation.

To overcome these problems, we propose an IRS-sharing system to achieve optimal radio wave propagation for multiple MNOs. Furthermore, we numerically evaluate the effectiveness of the IRS-sharing system and clarify the extent of the benefits provided by sharing IRSs. Through this analysis, we provide guidance on the potential use of IRS in the design of future 6G infrastructures.

\section{Sharing-IRS: Definition and Concept}
First, we introduce previously considered infrastructure-sharing strategies for BS and discuss their applicability to IRS-sharing systems. Following this, we propose our concept for an IRS-sharing system.

\subsection{Infrastructure Sharing Strategies and Applicability to IRS-Sharing}

\begin{figure*}[t]
    \centering
    \includegraphics[width=1\hsize]{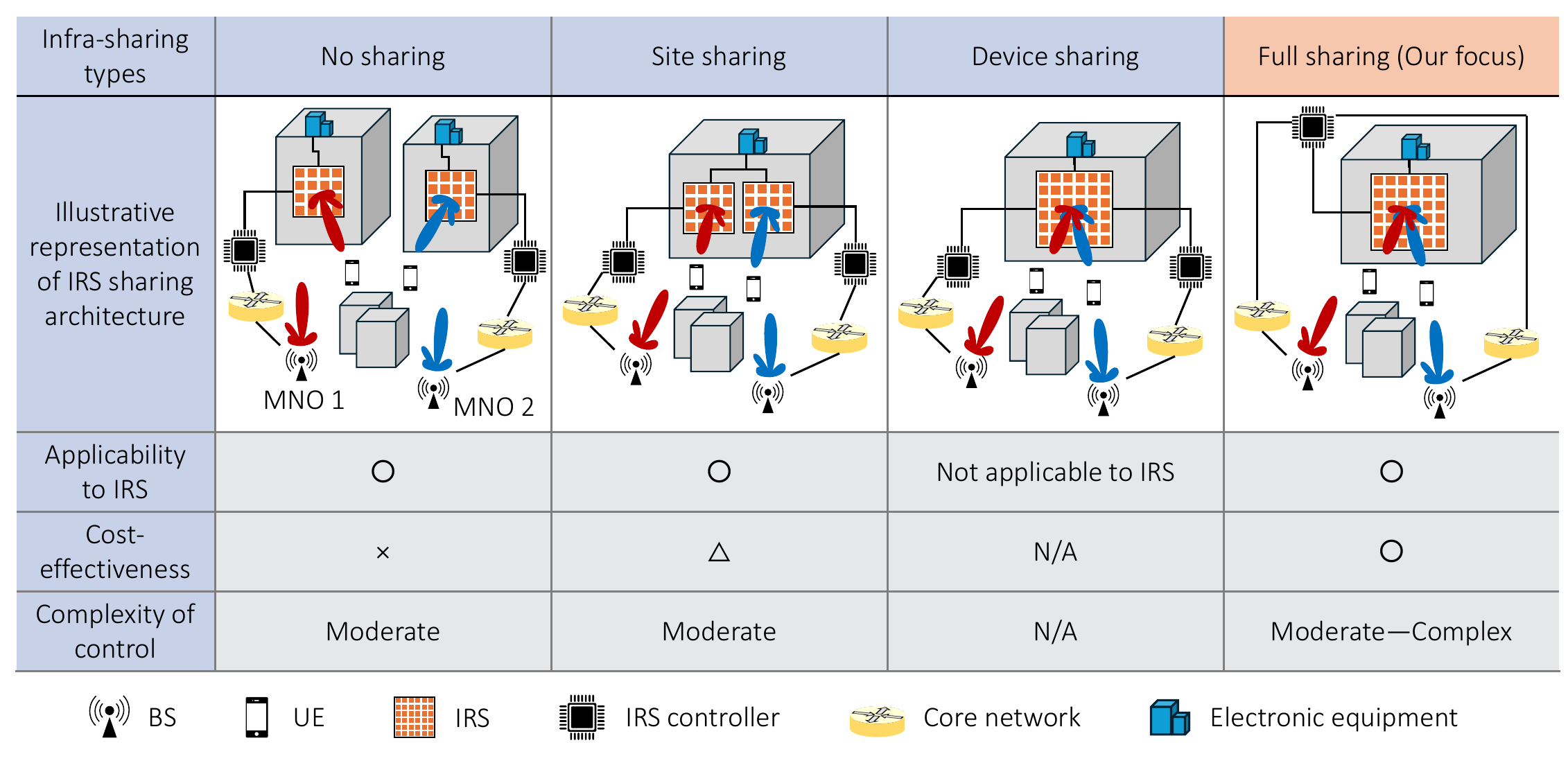}
    \caption{BS-sharing strategies and their applicability to IRS-sharing.}
    \label{fig:sharing_types}
\end{figure*} 

The open radio access network (O-RAN) alliance acknowledges the significance of establishing open-interface standards for communication equipment to achieve cost-effective and flexible performance enhancement. BS sharing among multiple systems has been investigated previously, with particular emphasis on sharing among different MNOs~\cite{infra_sharing}. First, we provide an overview of BS-sharing strategies and subsequently discuss their applicability to IRS sharing. In cellular communications, infrastructure sharing can be categorized into several strategies based on the scope of the shared components and the level of operational cooperation between MNOs (Fig.~\ref{fig:sharing_types}).

\subsubsection{Site Sharing} 
Site sharing is the most simple strategy in which the physical installation location of electronic equipment is shared~\cite{site_sharing}. It enables the integration and sharing of passive components such as power generation, conversion, and distribution. This approach presents challenges for IRS deployment despite being suitable for infrastructure such as base stations. IRSs must be installed on specific sites such as building facades or towers; hence, sharing sites is difficult. Therefore, the benefits of site sharing are limited in the context of IRS deployment.

\subsubsection{Device Sharing} 
Device sharing refers to sharing communication equipment such as radio heads and backhauls across multiple systems~\cite{Device_sharing}. By separating the core network or the control systems of the devices, each system can independently execute communication control, such as baseband processing and resource allocation, while sharing devices such as antennas. This separation is achieved through frequency division by isolating the signals of each system. In case of IRS, because IRS controls the phase of incident radio waves in the analog domain, independent control by each MNO is challenging. Unlike digital systems, in which signal separation can be achieved through frequency division or multiplexing, analog systems do not allow such separation. Therefore, active sharing of the IRS, where each MNO independently controls radio wave propagation, is not feasible.

\subsubsection{Full Sharing} 
Full sharing represents an arrangement in which all components, including passive components, electronic devices, and control systems, are fully shared. Full sharing is the most practical strategy for IRSs and enables the optimization of communication for all participating systems by sharing installation sites, devices, and control functions. Although this strategy maximizes the benefits of IRS sharing, it requires a high degree of cooperation and coordination among MNOs. In this study, we focus on full sharing and discuss the system architecture and control methods necessary for realizing this strategy.

\subsection{Our proposal: Concept of IRS Sharing}
\begin{figure}[t]
    \centering
    \includegraphics[width=1\hsize]{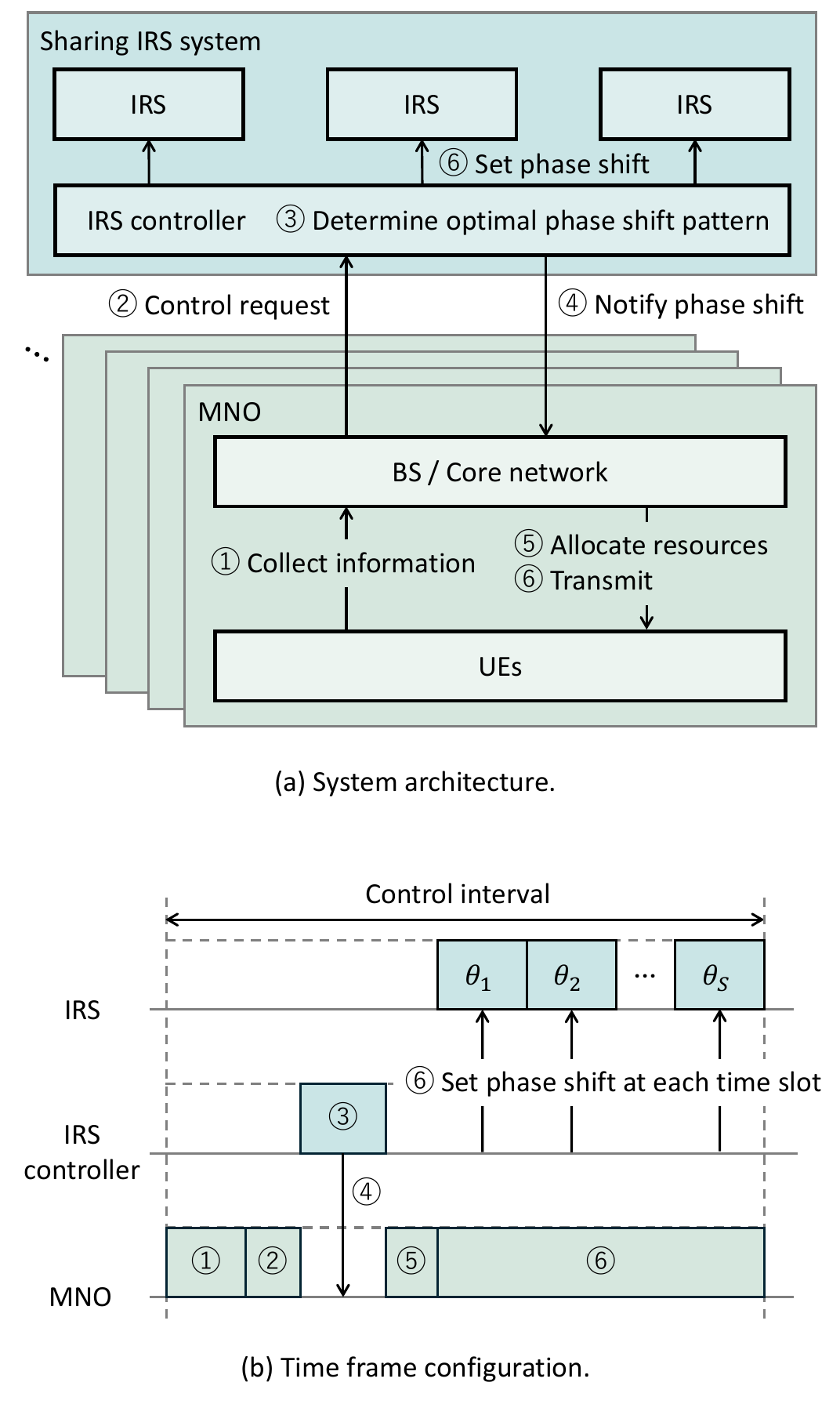}
    \caption{Proposed IRS-sharing system.}
    \label{fig:detailed_explanation}
\end{figure}

IRS sharing involves multiple MNOs collaborating to use a common IRS, as depicted in Fig.~\ref{fig:motivation_of_IRS_sharing}. In this system, several MNOs share the IRS, IRS controller, and electrical equipment required for their operation. These shared facilities operate independently of any single MNO's core network and position them as parallel entities. Therefore, shared infrastructure can be established either through joint investment by the MNOs using it, or through funding from a third party separate from the MNOs. 
The IRS controller manages all IRS units based on control requests received from each MNO. Although conventional IRS networks typically exchange control signals between a specific MNO’s core network and its IRSs, in an IRS-sharing scenario, the IRS controller receives control requests from multiple MNOs.
Fig.~\ref{fig:detailed_explanation} reveals a detailed explanation of the system architecture and timeframe configuration of the IRS sharing systems. 
The control requests information about the reflection control strategies that each MNO requires to improve communication quality within its network. The requests are transmitted through wired or wireless control links established between the MNOs and the IRS controller. The controller integrates these requests and adjusts the phase shifts of each IRS to minimize the areas in which the received signal strength falls below an acceptable level for any MNO.
Subsequently, the IRS controller communicates the determined scheduling of the IRS phase shifts to all MNOs through control links. The IRS phase shifts are then switched according to this schedule within a predefined control interval, enabling each MNO to synchronize its communication activities with its users accordingly. When a control interval expires, each MNO sends control requests to the IRS, initiating a new control interval.

The IRS controller ensures fairness in the communication environment for all MNOs by appropriately determining IRS control scheduling. 
To ensure fairness among MNOs, the controller uses a strategy to maximize the achievable rate of the UE with the lowest performance across all MNOs, namely, the UE that benefits the least from the IRS.
The IRS is a passive device; hence, a large aperture is required to share sufficient received signal power with users. Although this phenomenon increases the reflection gain of the IRS, it considerably narrows the beam width; thereby making it challenging to simultaneously deliver high received power to the users of multiple MNOs. To this end, the IRS control interval was categorized into multiple timeslots, and the reflection pattern was configured to ensure that all MNO users requiring IRS assistance could communicate throughout all timeslots. Additionally, the reflection pattern was optimized in each time slot to enhance the communication quality for as many MNOs as possible. 

\section{Advantages, Design Challenges, and Solutions}
In this section, we discuss the advantages of the IRS sharing system, the challenges associated with its implementation, and potential solutions to address these challenges.
\subsection{Advantage of Sharing-IRS}

\subsubsection{Flexibility}
Sharing an IRS among multiple MNOs enhances flexibility in network planning and deployment. Suitable locations for IRS installation are limited; hence, having separate IRS units for each MNO can lead to competition among MNOs. These locations can be used more efficiently by sharing an IRS, enabling an optimal deployment that maximizes the performance of each MNO. IRS sharing provides a flexible solution that can be adapted to the needs of multiple MNOs.

\subsubsection{Cost Saving}
By sharing IRSs among multiple communication systems, infrastructure costs can be distributed across multiple MNOs, considerably reducing capital and operational expenditures. This approach is advantageous in scenarios involving multiple operators and environments where small-scale communication systems using the same frequency band coexist. Deploying separate IRSs for each system can be prohibitively expensive. However, sharing a single IRS renders this technology feasible even in cases where individual systems have challenges in deploying their own IRS. IRSs are used to enhance signal coverage and quality for users obstructed by buildings or other obstacles. If each MNO were to independently use a dedicated IRS, in several cases, the IRS would be underutilized, resulting in resource wastage. Therefore, the idle time can be minimized by sharing the IRS across multiple systems, enabling an environmentally friendly and cost-efficient utilization of resources.

\subsubsection{Avoiding Inter-Operator Effects}
Through the integrated control of the IRS controller, MNOs can achieve indirect coordination by distributing synchronization signals without direct synchronization. This coordination is facilitated by the IRS controller, which distributes synchronization signals. This method enables preventing changes in the IRS phase shift during channel estimation or data transmission that could conflict with the intentions of the MNOs. Consequently, channel estimation errors can be reduced, and potential adverse effects on signal propagation among other MNOs can be minimized. Furthermore, the IRS controller ensures a fair communication environment for all MNOs by appropriately scheduling IRS control.

\subsection{Design Challenges of IRS-sharing Systems}
Although full sharing provides numerous advantages, the technique poses challenges, particularly in the joint design and control of the IRS reflection coefficients. Typically, IRS settings are optimized based on physical layer information such as channel state information (CSI). To enable the phase optimization of the IRS based on CSI, high-precision intersystem synchronization is required during channel estimation, optimization calculations, and channel information transfer. However, sharing such low-layer information with other MNOs is challenging from a confidentiality perspective, and achieving intersystem synchronization among various MNOs is difficult because of system independence. Therefore, novel algorithms and coordination mechanisms are needed to enable effective IRS sharing while simultaneously maintaining MNO independence.

\subsection{Promising Solutions}
In this subsection, we introduce several promising solutions to enable IRS sharing.

\subsubsection{Time-Division}
A simple method to share an IRS without sophisticated synchronization among MNOs is transferring IRS control to each MNO. In this approach, the IRS controller divides the control interval into several time slots corresponding to the number of MNOs and sets the IRS phase-shift pattern to reflect the control request of the MNO assigned to each time slot. This phenomenon enables the optimal radio wave propagation for the MNO assigned to the active timeslot. However, MNOs not currently assigned a timeslot can experience limited communication; hence, their overall communication capacity tends to decrease with the increase in the number of MNOs.

\subsubsection{Stand-Alone Switching}
Another promising approach involves periodically switching the IRS phase shifts. In this method, the IRS synchronizes with GPS or GNSS to periodically toggle multiple phase-shift patterns. Information related to the phase-shift switching schedule---specifically, the switching interval, the number of reflection patterns used in one switching cycle, and the start time of each cycle---is predetermined within the system and shared among all operators. In the first switching cycle, the MNO measures the channel quality information of the UE for each phase shift pattern. Based on this information, the MNO selects a pattern that improves the UE’s communication quality and allocates the corresponding time resources to the UE in the subsequent switching cycle. This method eliminates the necessity for the MNO to send control requests to an IRS controller. Furthermore, the MNO does not require the actual values of the phase shifts used, and sharing this information is not required. 
While the IRS switching schedule and initiation timing need to be pre-defined in the protocol, explicit coordination among MNOs is not required. Consequently, this method ensures complete independence between the sharing-IRS and the MNO.

\subsubsection{World Model-based Approach}
Another advanced approach involves optimizing the phase shifts of the IRS based on a world model~\cite{Hashida_sharing}. A world model is an internal virtual model that an agent uses to understand its surrounding environment, enabling it to plan and execute actions accordingly. Typically, the phase-shift optimization of an IRS focuses on maximizing the transmission capacity. To determine appropriate IRS phase shifts, gathering CSI and acquiring information about the wireless environment through an interactive process is essential. However, in scenarios involving shared IRSs, acquiring such information, including channel state information, becomes challenging, rendering the optimization process complex. To address this challenge, each MNO first develops a machine learning model that learns the changes in radio wave propagation within its own model when the IRS is set to a specific phase shift. 
The IRS controller is programmed to apply arbitrary phase shifts, with the phase shifts and the corresponding received power at various locations used as training data.
The machine learning model is trained using these phase shifts as inputs and the received power maps as outputs. 
The trained model is then shared with the IRS controller, which possesses the necessary computational capacity. These models function as world models for the IRS controller, helping determine appropriate IRS phase shifts\footnote{This approach is similar to federated learning; however, it differs in that each MNO independently trains its own model and utilizes it for shared IRS control. In federated learning, multiple agents typically collaborate to train a single global model. Conversely, this approach does not integrate the models trained by different MNOs into a single model. Instead, each trained model is used to determine the control strategy for the shared IRS.}. When a UE requires IRS assistance, the MNO sends a control request to the IRS controller specifying the required received power at each point. The IRS controller then searches for the optimal IRS phase-shift configuration to satisfy this control request using a world model. This search enables the realization of an optimal phase-shift pattern for all operators without requiring intensive collaboration.
In this approach, each MNO must develop its own machine learning model, which may increase the system complexity. However, this also enables more advanced phase shift configurations for the IRS.
The computational complexity of the proposed algorithm depends on the structure of the machine learning model used by the MNO. However, this computational complexity does not impact real-time communication since model training is executed asynchronously with communication.

\section{Numerical Analysis}
We conducted numerical analyses to verify the effectiveness of the IRS-sharing approach by comparing its performance with those of other benchmark methods. In scenarios where multiple MNOs share an IRS, ensuring fairness among MNOs is crucial. Particularly, any control method that operates with the advantages or disadvantages of a single MNO is unsuitable. Therefore, in this analysis, we evaluated the communication quality of the user with the lowest communication rate among those using IRS. This approach enabled us to determine whether the IRS sharing method could guarantee consistent communication quality for all MNOs involved.

\subsection{Simulation Settings}
We consider an environment in which there are $N$ MNOs; each MNO operates a BS and provides downlink communication to the UE through an IRS. IRS comprised $L \times L$ elemental configurations.
Each element of the IRS is arranged with a half-wavelength, and the area occupied by each element is $\frac{\lambda}{2} \times \frac{\lambda}{2}$. The transmission time is divided into $K$ time slots, and the IRSs can configure a distinct phase shift for each slot. 
The BS transmitted at a power of $30$~dBm; the transmission antenna gain was set to $20$~dBi, and the receiving antenna gain was $0$~dBi. Noise power was assumed to be $-80$~dBm, and the carrier frequency was $28$ GHz. The propagation channel followed the Rician fading model with a Rician factor of $0$~dB.
The IRS was positioned at the coordinates $(0, 0, 2~\text{m})$; its reflective surface was aligned with the $x$-$z$ plane. All MNO base stations shared the same site, located at $(10~\text{m}, 5~\text{m}, 5~\text{m})$. The users were randomly located within a square area at $(-5.5~\text{m}, 7.5~\text{m})$ with dimensions of $5~\text{m} \times 5~\text{m}$ in the horizontal plane. Their heights were randomly distributed between $1$ and $5$m.

In this analysis, we examined the achievable minimum rate in the proposed IRS sharing scheme under the assumption that the IRS and the MNO network are fully synchronized and perfect CSI is available, deriving the upper bound on the achievable performance.
To obtain CSI at each MNO, the IRS cycles through a set of predefined reflection patterns on a fixed schedule. MNOs synchronize their pilot signal transmissions with the IRS’s switching schedule by aligning their clocks using GPS or GNSS. Each MNO sends pilot signals that, when reflected by the IRS under different configurations, obtain different received pilot signals. With linearly independent reflection patterns, the unique pilot responses allow each MNO to accurately estimate the CSI.
However, the max-min optimization problem, which maximized the minimum achievable transmission rate among all users, involved finding the optimal phase shifts and was nonconvex, rendering obtaining a global optimum difficult. 
Thus, we developed an optimization algorithm based on a projected gradient descent to determine a near-optimal solution to address this problem.
In this algorithm, first, the phase shifts for each time slot were randomly initialized. Next, the transmission rates for each user under the current phase shift and the user $n^{*}$ with the lowest rate are identified. Next, we computed the gradient of the phase-shift matrix with respect to the transmission rate function of user $n^{*}$, updating the phase-shift matrix in a direction that improves the user rate. Following each update, the phase shifts were projected onto the unit circle, ensuring that each element of the phase-shift matrix had a magnitude of one. This process was iterated until the convergence criteria were satisfied. On convergence, the algorithm terminated and outputted the final phase shift matrix.
The computational complexity of this algorithm is $O(NKL^2)$ for each iteration\footnote{When the number of IRS elements is very large, clustering or grouping IRS elements can be applied to reduce the number of optimized variables. Furthermore, the matrix multiplication of the channel calculation and gradient updates for each IRS phase shift are inherently parallelizable. The computational load can be distributed across multiple processors or cores by leveraging multithreading or GPU acceleration, reducing processing time.}.

To demonstrate the effectiveness of the sharing approach, the performance of the following three benchmarks was evaluated under the same conditions as the sharing approach:
\begin{itemize}
    \item \textit{IRS sharing}: The control interval of the IRS was divided among the number of MNOs, with the IRS setting a different phase shift for each slot. The IRS phase shifts were optimized to maximize the minimum achievable rate across all users, representing the upper bound of achievable performance in the IRS sharing scenario.
    \item \textit{Time-division}: Similar to \textit{IRS sharing}, the IRS control interval was divided according to the number of MNOs, with various phase shifts configured for each slot. However, in each time slot, the IRS phase shift was optimized exclusively for the assigned MNO.  
    \item \textit{No sharing}: In this approach, the IRS was not shared among MNOs. The IRS was divided into sub-surfaces of size $\lfloor \sqrt{L/N} \rfloor \times \lfloor \sqrt{L/N} \rfloor$, and each sub-surface was allocated to a specific MNO, who could then optimize the phase shifts of their assigned IRS sub-surface.
    \item \textit{Random phase shift}: The phase shift for each IRS element was randomly set within the range $(0, 2\pi)$.  
\end{itemize}

\subsection{Results}
\begin{figure}[t]
    \centering
    \includegraphics[width=1\hsize]{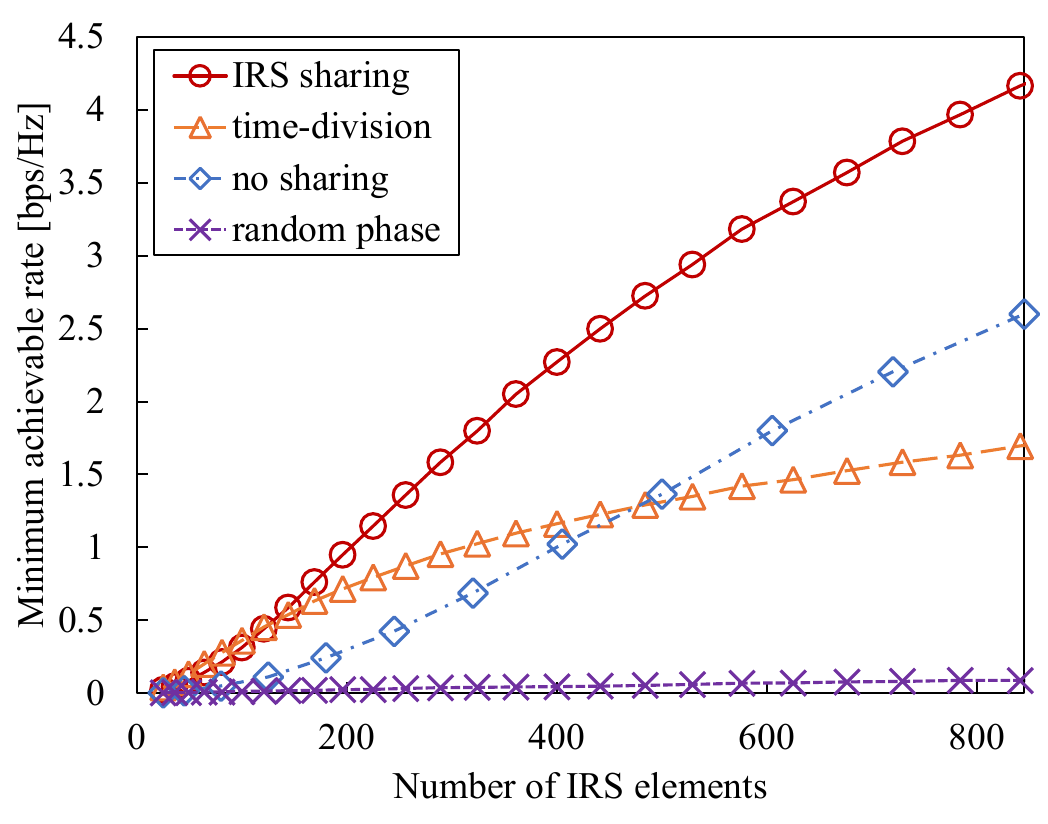}
    \caption{Numerical result of the minimum achievable rate versus the number of IRS elements.}
    \label{fig:vs_element}
\end{figure}

Fig.~\ref{fig:vs_element} depicts the minimum achievable rate as a function of the number of IRS elements $L^2$ under the presence of $5$ MNOs. The graph reveals that when the number of IRS elements is lower than $200$, the performance of the IRS-sharing approach is comparable to that of the time division approach. However, when the number of IRS elements exceeds $200$, the improvement in performance provided by the time division approach becomes limited; IRS sharing consistently achieves the best performance. In the time division scheme, the beam is focused on the UE of a specific MNO during its designated slot, resulting in periods when other UEs cannot communicate. Conversely, the IRS-sharing scheme determines the IRS control strategy by considering fairness among all UEs, enabling simultaneous support for multiple users. Additionally, when the total number of IRS elements is small, the number of elements that a single MNO can control in the no-sharing scheme is limited, which restricts the reflected signal power available to UEs and results in a lower achievable rate. However, with an increasing number of IRS elements, sufficient signal power can be consistently reflected to UEs, which results in a better performance than time division. For random phase shifts, even with numerous elements, communication is not feasible, highlighting the necessity of optimizing phase shifts, specifically for UEs.

\begin{figure}[t]
    \centering
    \includegraphics[width=1\hsize]{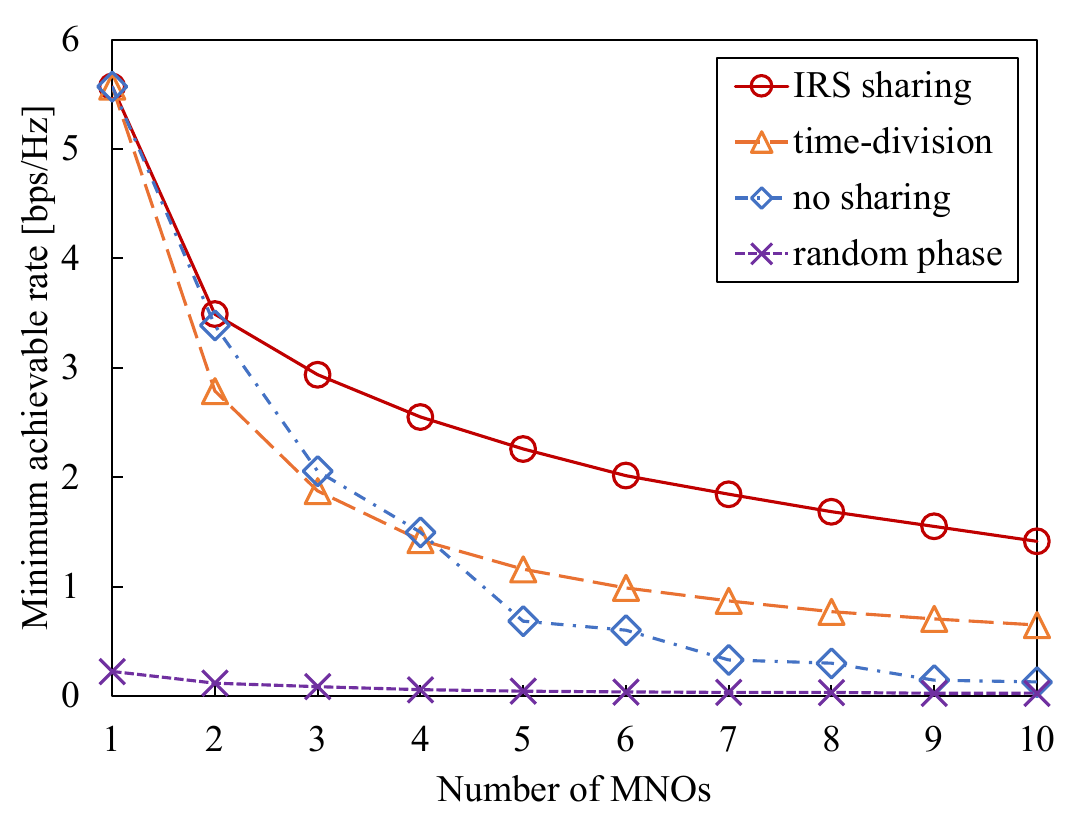}
    \caption{Numerical result of the minimum achievable rate vs. the number of MNOs.}
    \label{fig:vs_MNO}
\end{figure}

Fig.~\ref{fig:vs_MNO} illustrates the minimum achievable rate as a function of the number of MNOs for $400$ IRS elements. When the number of MNOs was one, the IRS sharing, time division, and no-sharing configurations yielded identical performances because they operated under the same conditions. However, as the number of MNOs increased, the performances of both the time-division and no-sharing configurations deteriorated considerably. In the case of time division, the communication time is divided among the MNOs, which reduces the duration for which the IRS can reflect adequate power for the UEs of any single MNO, resulting in performance degradation with the increase in the number of MNOs. Similarly, in the no-sharing case, the total number of IRS elements is limited; as the number of MNOs increased, the number of IRS elements available to each MNO decreased, which resulted in insufficient power reflection for each UE.
The IRS-sharing configuration also experienced a performance decline with an increasing number of MNOs due to the growing difficulty in balancing the demands of multiple MNOs. However, this degradation was more gradual than that in the other methods. Therefore, the results demonstrated that IRS sharing remains effective in multi-MNO scenarios, indicating its potential as a viable solution to accommodate multiple MNOs.

\section{Open Issues}
In addition to design problems related to IRS sharing, other significant open problems should be investigated. In the following section, we outline the extendable directions to motivate future work.

\subsection{Deployment of Sharing IRS} Determining the optimal placement of an IRS is crucial for effective shared IRS deployment. Because of the analogous nature of IRSs, the phase shifts cannot be configured independently for each frequency. Consequently, the phase-shift configuration’s capability to simultaneously cover areas of weak signal power for users of multiple MNOs is highly dependent on IRS placement. The optimal deployment strategy for a shared IRS could differ considerably from conventional approaches used for a single MNO, which should be studied in the future.

\subsection{Avoiding Malicious Shared IRS Control} The phase shift configuration strategy for a shared IRS should be secured to prevent hacking by malicious MNOs. For example, an MNO can send manipulated control requests to an IRS controller, resulting in phase shifts that are advantageous or disadvantageous for specific MNOs. Therefore, developing robust control methods is critical for defending against disruptive interference.

\subsection{Multi-/Wide-band Intelligent Reflecting Surface} Typically, each MNO operates on different frequency channels. IRSs generally function within a specified operating bandwidth---approximately $\pm 10$~\% of the design frequency---covering the frequency channels used by each MNO operating in the same band. However, the frequency characteristics of IRS elements are not uniform and exhibit distortion across various frequencies, which can affect fairness among MNOs. Therefore, investigating the control methods for shared IRSs that consider these frequency characteristics is crucial. Furthermore, the development of IRSs that can perform multi-band or wideband operations is a promising approach.

\section{Conclusions}
This study focused on IRS operations in environments with multiple MNOs and proposed a novel framework for sharing IRS among these MNOs. 
We proposed sharing IRSs, a cost-effective approach that enables multiple MNOs to use common IRSs. The proposed approach ensures consensus on IRS control requests from each MNO, avoiding unintended modifications to radio wave propagation by the IRS that could adversely affect any MNO and maximize the benefits derived from the IRS. Numerical evaluations demonstrate that IRS sharing contributes to considerable improvement in radio wave propagation environments across multiple MNOs, proving effective in multi-MNO scenarios. Consequently, this framework presents an efficient method for leveraging the IRS and supporting the development of sustainable next-generation networks.

\section*{Acknowledgment}
A portion of the results presented in this paper was obtained through research commissioned by JSPS KAKENHI Grant Number 24K23850.

\bibliographystyle{IEEEtran}
\bibliography{reference} 

\begin{IEEEbiography}[{\includegraphics[width=1in,height=1.25in,clip,keepaspectratio]{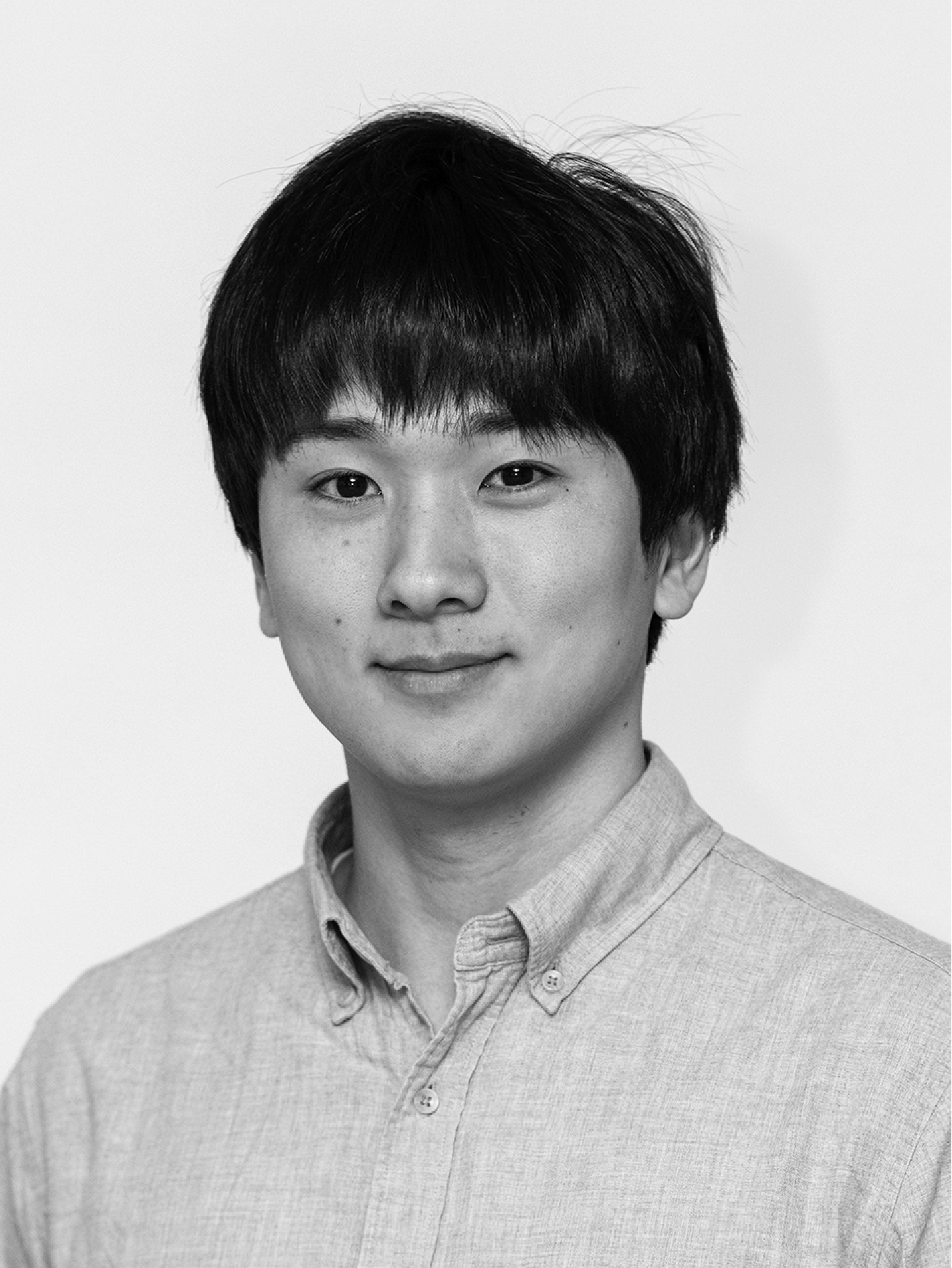}}]
{Hiroaki Hashida}.
(M’24) is an Assistant Professor with the Frontier Research Institute for Interdisciplinary Sciences, Tohoku University, Japan. He received his Ph.D. in 2024 from Tohoku University, Sendai, Japan. He was a recipient of the Presidential Award for Outstanding Students from Tohoku University and Ikushi Prize from the Japan Society for the Promotion of Science in 2024. His research interests include wireless communication networks, and intelligent surface-aided wireless communication systems. He is a member of the IEEE and Institute of Electronics, Information, and Communication Engineers.
\end{IEEEbiography}

\begin{IEEEbiography}[{\includegraphics[width=1in,height=1.25in,clip,keepaspectratio]{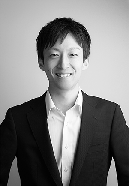}}]
{Yuichi Kawamoto}
(M'16) is a full professor at the Graduate School of Information Sciences (GSIS) at Tohoku University, Japan. He received his M.S. degree in 2013 and completed his Ph.D. in Information Science in 2016 from Tohoku University, Japan. He has published more than 60 peer-reviewed papers, including several high-quality publications in prestigious IEEE journals and conferences. Notably, he received the best paper awards at several international conferences, including IEEE flagship events, such as the IEEE Global Communications Conference in 2013 (GLOBECOM’13), the IEEE Wireless Communications and Networking Conference in 2014 (WCNC’14), and the IEEE International Conference on Communications in 2018 (ICC’18). In addition, he was a recipient of the Dean’s and President’s Awards from Tohoku University in 2016. His research interests cover a wide range of areas, including satellite communications, unmanned aircraft system networks, wireless and mobile networks, ad hoc and sensor networks, green networking, and network security. Moreover, he is a member of the IEEE and the Institute of Electronics, Information, and Communication Engineers (IEICE).
\end{IEEEbiography}

\begin{IEEEbiography}[{\includegraphics[width=1in,height=1.25in,clip,keepaspectratio]{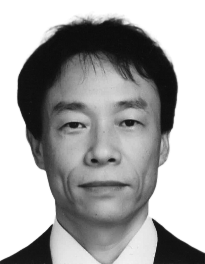}}]
{Nei Kato}
(F'13) is a full professor and dean at the Graduate School of Information Sciences, Tohoku University. He has studied computer networking, wireless mobile communications, satellite communications, ad hoc \& sensor and mesh networks, UAV networks, AI, IoT, and Big Data. He has published more than 500 papers in prestigious peer-reviewed journals and conferences. He is the Editor-in-Chief of the IEEE Internet of Things Journal and the Director of the Magazine of the IEEE Communications Society. He is a fellow at the Engineering Academy of Japan, IEEE, and IEICE.
\end{IEEEbiography}

\end{document}